\documentclass[12pt]{article}
\usepackage{amsmath}
\usepackage{amssymb}

\newcommand{\R}{\ensuremath{{\mathbb R}}}

\newcommand{\Z}{\ensuremath{{\mathbb Z}}}

\begin{document}
\thispagestyle{empty}
\vspace*{30pt}\begin{center}
{\LARGE\textbf{Non-Commutative Time, the\\ \vspace{7pt}Quantum Hall Effect and Twistor \\ \vspace{7pt}Theory}}\bigskip\vspace{10pt}\\
{\large\textmd{
Dana Mihai, George Sparling and Philip Tillman\\\vspace{10pt}Laboratory of Axiomatics\\\vspace{5pt}
University of Pittsburgh\\\vspace{10pt}Pittsburgh, Pennsylvania, 15260, USA}}\end{center} \vspace{40pt}
\begin{center}\textbf{Abstract}\end{center}
It is proposed that Maxwell theory, with a topological term, in four non-commutative dimensions, where the co-ordinates obey the Heisenberg algebra,  is an umbrella theory for the description of the two-dimensional Quantum Hall Effect (following the fluidic approach of Susskind and Polychronakos), the twistor analyses of self-dual gravity, solitons and instantons and the twistor theory of isolated horizons and black holes in space-time.  Applied to the Quantum Hall case, the underlying metric is Lorentzian: two canonically conjugate co-ordinates are spatial, the other two co-ordinates naturally are null and represent time and a conjugate energy variable.   \\\\
\begin{center}
The Laboratory of Axiomatics acknowledges support from a NATO grant: PST.CLG.978984.
\end{center}

\clearpage\mbox{}\setcounter{page}{1}

\section*{}
The issue of whether or not a mathematical model of a physical system correctly incorporates all the relevant dynamics of the system is not always easy to decide.  For example in supersymmetric systems, formulated in a Lorentzian framework, with at least one unbroken supersymmetry, one could argue that the dynamics is frozen out by the supersymmetry, which usually generates under (anti)-commutation a null or timelike symmetry.  In principle this can be remedied by going to a two-time formalism, with the symmetry removing one time variable and leaving a proper dynamical theory.    The theory would then be a quantum generalization of the approach of Charles Fefferman and Robin Graham, who, in order to study aspects of a Lorentzian system with group $\mathbb{O}(n, 1, \mathbb{R})$ would analyze an associated system with group $\mathbb{O}(n+1, 2, \mathbb{R})$, equipped with an appropriate symmetry [1,2].   In such an approach, the two "time" parameters are classical.  In this work we will argue that to describe the dynamics of the Quantum Hall Effect in two dimensions, it is appropriate to use a four-dimensional formalism, consisting of two mutually canonically conjugate spatial variables, one time variable and one energy variable, canonically conjugate to the time  [3-9].  \\\\
The origin of the present work lies in the observation due to the second author that the four-dimensional fermionic gauge quantum liquids constructed by Shou-Cheng Zhang and Jiang-Ping Hu are best understood as ordinary (non-gauged) quantum fermionic liquids in six-dimensional twistor space [10-26].   As such it behooves us to try to understand the dynamics of such fluids.  The twistor space description of the fluid in six dimensions is directly analogous to that of Duncan Haldane for the fermionic fluids corresponding to the  Quantum Hall Effect in two dimensions [6, 25].  Accordingly, as a step to a better understanding of the twistor fluid, we have returned to the two-dimensional case to see if we can comprehend the dynamics there.  One key  point here is that any results are in principle experimentally testable, in the laboratory, so should give a secure foundation on which to build the full revolutionary quantum twistor theory of space-time.  \eject\noindent
Perhaps the leading approach to the theory of the Quantum Hall Effect is that of Leonard Susskind and Alexios Polychronakos [27-31].  Susskind, in a very deep and difficult analysis of the Quantum Hall Effect makes the critical observation that the granular nature of the quantum fluid can be reflected first in a theory invariant under the group of area preserving diffeomorphisms of the plane and then in  a theory based on its quantum analogue: the Heisenberg algebra for one degree of freedom, in its Moyal representation, with its symmetry group, the group of all unitary automorphisms of the algebra.  He then adds in a time variable and constructs a non-commutative generalization of Chern-Simons theory, based on the Moyal description of the enveloping algebra of the Heisenberg algebra, which is supposed to give the desired dynamical theory.  Polychronakos then improves the formalism, basing it on his $\mathcal{D}$-operator (a combination of the generalized exterior derivative and the gauge potential), here called $\mathcal{Q}$.\\\\
Remarkably, the group of two-dimensional area preserving diffeomorphisms and its Moyal generalization occur naturally in at least two other important areas of physics:  first in the context of  the search for a unified theory of soliton equations, particularly in the description of the Kadomtsev-Petviashvili equation; second in the description by twistor theory of the solutions of the self-dual Einstein vacuum equations [32-38].  It is our basic contention that this is no accident: that there is a single theory which encompasses all three aspects of this group.  To see the emergence of our viewpoint, consider Hamiltonian evolution in a two-dimensional phase space.  With respect to canonical co-ordinates $p$ and $q$, Hamilton's equations for a conservative system are: $\frac{dq}{dt} = \frac{\partial H}{\partial p}$,  $\frac{dp}{dt} = -\frac{\partial H}{\partial q}$,  where $t$ is the physical time parameter and the Hamiltonian $H(p, q)$ is a function of the phase-space variables $p$ and $q$ only.  The evolution preserves the symplectic two-form  $dp\wedge dq$, so the dynamics proceeds by means of area preserving diffeomorphisms of the $(p, q)$ space (the point of view exploited particularly by Liouville).  However in the case that the system is not conservative, so that the Hamiltonian depends on $p$, $q$ and $t$, it is known that the correct way to describe the system is by means of the contact form $pdq - Kdt$ and its associated symplectic form $dp\wedge  dq - dK\wedge dt$, where the Hamitonian  is promoted to a new variable $K$ conjugate to the time variable.  Then the new Hamiltonian in four dimensions is the function $K - H(p, q, t)$. 
\eject\noindent
Now it is a fact that the twistor description of the self-dual Einstein vacuum equations uses a time-dependent Hamiltonian formalism. In the language of the good-cut equation of Ted Newman, using the standard two-component spinor formalism of Sir Roger Penrose, the propagation equation is [39,40]:
\[\frac{\partial \omega^A}{\partial \overline{\pi}_B} = \overline{\pi}^A\overline{\pi}^B\sigma(-i\omega^C\overline{\pi}_C, \pi_{D'}, \overline{\pi}_D).\]
Here $\sigma(u, \pi_{D'}, \overline{\pi}_D)$ is a given real analytic function which obeys the homogeneity requirement $\sigma(\lambda\overline{\lambda} u, \lambda\pi_{D'}, \overline{\lambda}\overline{\pi}_D) = \lambda(\overline{\lambda})^{-3} \sigma(u, \pi_{D'}, \overline{\pi}_D)$, for $\lambda$ any non-zero complex number.  The desired solution $\omega^A(\pi_{B'},\overline{\pi}_{B})$ is required to obey the homogeneity condition $ \omega^A(\lambda\pi_{B'}, \overline{\lambda}\overline{\pi}_B) = \lambda\omega^A(\pi_{B'}, \overline{\pi}_B)$, again for $\lambda$ any non-zero complex number.   If that solution is globally defined, then it represents a point of the space-time.  For example the global solutions in the case that $\sigma = 0$ are $\omega^A = ix^{AA'}\pi_{A'}$, where $x^a$ is constant and then the associated space is the complexification of Minkowski spacetime.\\\\  Physically the function $\sigma$ represents the radiation data at null infinity of a real space-time [41].
A prototypical example with $\sigma$ non-zero is the case that $\sigma = \frac{f(\pi_{A'})}{(u + it^{a}\pi_{A'}\overline{\pi}_A)^3}$, with $f(\pi_{A'})$ a holomorphic homogeneous polynomial of degree four: $f(\pi_{A'}) = \phi^{A'B'C'D'}\pi_{A'}\pi_{B'}\pi_{C'}\pi_{D'}$, where $\phi^{A'B'C'D'}$ is a constant totally symmetric spinor.  Also $t^a$ is a timelike real vector.  This corresponds to a quadrupole radiation field [42]. From the vantage of the Quantum Hall Effect, we regard $\overline{\pi}_{A}$ as being the (complex) time parameter and we can think metaphorically of $\omega^A$ as describing the position of an "electron" as time evolves.   Then the points of the space-time are described by those "electrons" that evolve globally over all time (the Riemann sphere represented by the spinor $\pi_{A'}$).  Since $\sigma$ depends explicitly on the variable $\overline{\pi}_{A}$, the corresponding Hamiltonian is "time" dependent.
\\\\
The analogy to the Quantum Hall theory becomes much closer if one considers the analogue of the Newman-Penrose theory for the case of  real self-dual spacetimes with a real metric of ultra-hyperbolic signature $(2,2)$.  In this case the spinors $\pi_{A'}$ and $\overline{\pi}_A$ become independent real spinors (with symmetry group $\mathbb{SL}(2, \mathbb{R})$).    Then $\overline{\pi}_A$ represents a "real" time and we have a one-parameter ensemble of Quantum Hall-like systems, parametrized by the projective parameter $\pi_{A'}$.  The ultra-hyperbolic signature was shown by the second author and Lionel Mason to be the correct one for describing integrable systems, solitons and the three-dimensional CT scan [43,44].\eject\noindent
In the twistor case, it is the two-form $d\omega^A\wedge d\omega_A$, at fixed $\pi_{A'}$, that is preserved by the dynamics and, as first shown brilliantly by Penrose,  the fact that this two-form is preserved encodes the information that the space-time associated to the twistor space is a (complex) solution of the Einstein vacuum equations [35].\\\\
The first part of our proposal now is this: we expect the general dynamics of Quantum Hall systems to be governed by a time-dependent Hamiltonian theory, exactly as in the twistor theory.  Both theories then become aspects of a single theory.  Accordingly the relevant phase space should be four-dimensional.  At the quantum level we will have a Heisenberg algebra corresponding to a system of two quantum degrees of freedom, rather than the single degree of freedom used in the theory of Susskind.  This entails introducing an "energy" operator canonically conjugate to the physical time.  The full quantum commutator algebra then reads:
\[ x^j x^k - x^k x^j = i\omega^{jk}.\]
Here $i^2 = -1$ and the lower-case Latin indices range from one to four.  The operators $x^j$ are understood to be Hermitian and then the c-number tensor $\omega^{jk}$ is a real skew non-degenerate (constant) symplectic form in four-dimensions. \\\\
What then is the theory?  It must include the theories of Susskind and Polychronakos in the limit of time-independent Hamiltonians, giving us an obvious candidate, which we will adopt as our model: non-commutative Maxwell theory with a topological term.  So our proposed action $\mathcal{S}$ reads as follows:
\[ \mathcal{S} = \lambda \int \mathcal{Q}^4 + \int \mathcal{Q}^2*\mathcal{Q}^2.\]
Here $\lambda $ is a coupling constant, which plays the role analogous to the magnetic field strength in the Susskind theory.  The odd operator $\mathcal{Q}$ contains the information of both the exterior derivative and the Maxwell potential one-form, classically called $A$. The even operator $\mathcal{Q}^2$ is then the analogue of the Maxwell field two-form, classically called $F = dA$, so the first term in the Lagrangian generalizes the topological term $\int F\wedge F$ of ordinary Maxwell Theory.  The second term generalizes the term $\int F\wedge (*F)$, the standard Maxwell Lagrangian, where $*$ is the duality operator on two-forms.  So to fix the theory we need an underlying metric (or conformal structure) in order to be able to define the duality operator.  \eject\noindent Although ultimately one would use an axiomatic approach to the duality operator, for the present we will associate the duality operator to an ordinary constant symmetric invertible bilinear form $g_{jk}$.  Then the key question is the signature of the metric $g_{jk}$: Lorentzian (hyperbolic), Euclidean (elliptic), or ultra-hyperbolic?  In the Lorentzian case, the $*$ operator has square the negative of the identity.  In the other two cases, the $*$ operator has square the identity.  We identify the latter two cases with variants of the twistor theory associated with instantons in the Euclidean case and soliton theory in the ultra-hyperbolic case.  In each of these cases, there are two preferred values of the parameter $\lambda $, namely $\lambda = \pm 1$, for which the action is either self-dual or anti-self-dual (note that the totally skew part of the tensor $\omega^{pq}\omega^{rs}$ gives a preferred orientation, so these cases are distinct).  The metric in the ultra-hyperbolic case coincides with the metric of Fefferman-Graham type canonically associated to a second-order differential equation, as discovered last year by the second author and Pawel Nurowski.  The Lorentzian case is then left for the description of the Quantum Hall fluid.   \\\\
Now given a real vector space equipped with a symplectic form $\omega^{jk}$ and a metric $g_{jk}$ there is canonically associated an invertible endomorphism $h_{i}^{j} = g_{ik}\omega^{jk}$ (using the Einstein summation convention).   The structure is then classified by the eigen-values of the matrix $h$.  Since the $*$ operator of our Lagrangian only depends on $g_{ij}$ up to scale, we may normalize by requiring that $\det(h) = \pm 1$.   Since the symplectic form is nondegenerate its determinant (being the square of its Pfaffian) is positive, so $\det(h)$ and $\det(g)$ have the same sign. Accordingly the sign of $\epsilon$ separates the standard twistor theories from the Quantum Hall theory: $\epsilon = 1$ for the soliton and instanton cases and $\epsilon = - 1$ for the Quantum Hall case.  In the Quantum Hall case we find that there always exists a natural co-ordinate basis ${x, y, t, E}$ with the following commutators: $xy - yx = ia$ and $tE - Et = ib$, where $x$ and $y$ are the spatial co-ordinates, $t$ is time and $E$ is energy and the pairs $\{t, E\}$ and  $\{x, y\}$ mutually commute and such that the metrical "squared interval" in this co-ordinates system is $Et + tE - x^2 - y^2$:  i.e. $E$ and $t$ are null co-ordinates and $x$ and $y$ are spatial co-ordinates.  Then we obtain the correct Euclidean (Newtonian) geometry on the level surfaces of the  $E$ co-ordinate.\eject\noindent
The second part of our proposal provides a sytematic method of integration in the non-commutative context.  We need to go beyond the Moyal expansion method of Susskind and Polychronakos, because, when that method is applied to the Maxwell or Chern-Simons Lagrangians, all the corrections arising from non-commutativity of the co-ordinates form exact derivatives and integrate out.\\\\
Here we consider only generalizations of ordinary integration, tracing back to the work of Archimedes and do not embrace the Berezin theory of fermionic integration, which requires a separate discussion and is not yet needed (although at a later stage, one could introduce fermions into our Lagrangian, to describe extended structures such as vortices).  Our point of view  (which is compatible with the approaches of Alain Connes and of Joachim Cuntz and Daniel Quillen) is that integration is essentially a sum [45,46].  In ordinary real integration we sum real numbers and the sum is also a real number.  This extends easily to complex integration, where the result is usually a complex number.  The general quantity we wish to integrate lies in a $\Z_2$-graded associative algebra, the quotient of a completion of the enveloping algebra of a  $\Z_2$-graded Lie algebra by a two-sided ideal (here $\Z_2$ is the group of integers modulo $2$).  To integrate we need only to know how to pull back this quantity to the standard arena of differential forms.  Then these differential forms will take values in the original $\Z_2$-algebra, but are integrated in a completely standard fashion.  In practice, usually we can rely also on the fact that we have available an appropriate represention of our algebra on some Hilbert space.   Then if the matrix elements of the pull-back form are taken with respect to states in the Hilbert space, the remaining integral is precisely that of a complex differential form.  The final result of the calculations gives an operator in the Hilbert space.  From this angle there is no specific need in general for a preferred exterior derivative operator as such in the original algebra.  
\\\\
We illustrate by considering the following integral:
\[ \int F(x,y) dxdy.\]
Here $x$ and $y$ are to obey the Heisenberg commutation relations $xy - yx = it$, with $t$ a real c-number.  The operator $d$ is defined as the ($\Z_2$-graded) commutator with the operator $\mathcal{Q}$ which itself obeys the relation $it\mathcal{Q} = xdy - ydx$, or equivalently $2it\mathcal{Q} = x\mathcal{Q}y - y\mathcal{Q}x$.
\eject\noindent We parametrize the integral with unitary transformations $U = e^{ipx_0 + iqy_0}$, which we map into the Heisenberg algebra by the formulas: $x = Ux_0U^{-1}$ and $y = Uy_0U^{-1}$.  Here $x_0$ and $y_0$ are ``base-point'' operators, obeying the Heisenberg algebra.  Also $p$ and $q$ are real parameters, which together range over the whole real plane, $\mathbb{R}^2$. Then the integral becomes:
\[ \Phi(x_0, y_0) = t^2 \int_{\mathbb{R}^2} F(x_0 - tq, y_0 + tp) dp dq.\]
Note that the operators $y_0 - i\partial_q$ and $x_0 - i\partial_p$ commute with the integrand, so, after integrating by parts, assuming that the function $F$ falls off at infinity appropriately, we find that $\Phi(x_0, y_0)$ is a central element, so in an irreducible unitary representation is a multiple of the identity operator.  That multiple then gives the desired numerical value to the integral.
The fundamental example is as follows:
\[ \pi_t = \int e^{-x^2 - y^2} dx dy.\]
The classical version of this integral, when appropriately generalized to infinite dimensions lies at the basis of bosonic quantum field theory. For the quantum case, where $x$ and $y$ obey the algebra $xy - yx = it$, where $t$ is a positive real c-number, the pullback integral is as follows: 
\[ \pi_t = \int e^{- x^{2}- y^{2}}dxdy = ite^{-t} \int_{\mathbb{C}}
e^{-2t (a^{\dag }-\overline{\lambda})(a- \lambda)) }d\overline{\lambda}d\lambda. \]
Here $\lambda = \sqrt{\frac{t}{2}}(p - iq)$.  Also the operators $a=\frac{x_0+iy_0}{\sqrt{2t }}$ and $a^{\dagger }=\frac{x_0-iy_0}{\sqrt{2t }}$ obey the harmonic oscillator algebra $aa^{\dagger} - a^{\dagger} a = 1$. We evaluate the integral by taking matrix elements with respect to coherent states for the operator $a$.  For any complex numbers $z$ and $w$, put $|z\hspace{-3pt}>= e^{za^{\dag }}|0\hspace{-3pt}>$ and $<\hspace{-3pt}\overline{w}|=<\hspace{-3pt}0| e^{\overline{w}a}$, where the vacuum state $|0\hspace{-3pt}>$ and its conjugate $<\hspace{-3pt}0|$ obey the relations $a|0\hspace{-3pt}>= 0$ and $<\hspace{-3pt}0|a^{\dag }=0$.  We observe that $a|z\hspace{-3pt}>=z|z\hspace{-3pt}>$,  $<\hspace{-3pt}\overline{w}
|a^{\dag }=<\hspace{-3pt}\overline{w}|\overline{w}$ and $<\hspace{-3pt}\overline{w}|z\hspace{-3pt}> = e^{\overline{w}z}$. Also $\left\{ |z\hspace{-3pt}>: z\in \mathbb{C}\right\} $ is an
over-complete set of basis vectors for the Hilbert representation space.  After normal ordering the integrand, we find:
\[ <\hspace{-3pt}\overline{w}|\pi_{t}|z\hspace{-3pt}> = i t e^{-t}<\hspace{-3pt}\overline{w}|z\hspace{-3pt}>\int_{\mathbb{C}}e^{( e^{-2 t}-1) (\overline{w}-
\overline{\lambda})(z-\lambda)}d\overline{\lambda}d\lambda \]
\[ = 2t e^{-t}<\hspace{-3pt}\overline{w}|z\hspace{-3pt}>\int_0^\infty \int_0^{2\pi}e^{( e^{-2 t}-1)
r^{2}}rd\theta dr=\frac{\pi t}{\sinh(t)}<\hspace{-3pt}\overline{w}|z\hspace{-3pt}>.\]
Hence $\pi_t$ is the real number $\frac{\pi t }{\sinh(t) }$ times the identity operator.  In the limit as $t\rightarrow 0$, we recover the standard classical value of $\pi$.
\eject\noindent
Summarizing, we have put forward a non-commutative Maxwell theory in four dimensions as an umbrella theory for the Quantum Hall effect, the theory of integrable systems and self-dual twistor theory. Presumably, when the Quantum Hall theory is constructed for a bounded spatial domain, the dynamics of the edge-states will then be governed by a three-dimensional theory, related to Chern-Simons theory and string theory [5,9].  This is currently under investigation. \\\\Plainly the present theory is capable of wide generalization. Of particular interest, perhaps, are the implications for the thermodynamics of systems with a non-commutative time operator.  We highlight one particular area, that of black holes. The isolated horizons, studied by Abhay Ashtekar and his collaborators, have the property that they are shear-free: this entails that the self-dual twistor theory is relevant for their description [47].  One might conjecture that, if the present theory is applied suitably, then the known thermodynamic and quantum properties of black holes, deduced by Penrose, Jacob Bekenstein, Stephen Hawking and others, would emerge naturally [48-53].


\newpage\begin{thebibliography}{30}
\bibitem{Fe1} Charles Fefferman and C. Robin Graham, \hspace{3pt}\emph{Conformal Invariants}, \hspace{3pt}in \emph{Elie Cartan et les Mathematiques d'Aujourdhui},\hspace{3pt} Asterisque, 95, 1985.
\bibitem{Gr1} C. Robin Graham, Richard Jenne, Lionel J. Mason and George A.J. Sparling, \hspace{3pt}\emph{Conformally invariant
powers of the Laplacian}, \hspace{3pt}Journal of the London Mathematical Society, 46(3), 557-565, 1992.
\bibitem{Kl1} Klaus von Klitzing, Gerhard Dorda and Michael Pepper,\hspace{3pt}
\emph{New Method for High-Accuracy Determination of the Fine-Structure Constant Based on Quantized Hall Resistance},\hspace{3pt}
Physical Review Letters, 45, 494-497, 1980. 
\bibitem{Ts1} Daniel C. Tsui, Horst L. Stormer and Arthur C. Gossard, \hspace{3pt}\emph{ Two-Dimensional Magnetotransport in the Extreme Quantum Limit}, \hspace{3pt} Physical Review Letters, 48, 1559-1562, 1982.
\bibitem{Pr1}Richard E. Prange and Steven M. Girvin,\hspace{3pt} \emph{The Quantum Hall Effect}, \hspace{3pt}Springer Verlag,
Berlin, Germany, 1990. 
\bibitem{Ha1}F. Duncan M. Haldane, \hspace{3pt}\emph{Fractional Quantization of the Hall Effect: A Hierarchy of Incompressible Quantum Fluid States},
\hspace{3pt}Physical Review Letters, 51, 605-608, 1983.
\bibitem{La1}Robert B. Laughlin, \hspace{3pt}\emph{Anomalous Quantum Hall Effect: An Incompressible Quantum Fluid with Fractionally Charged
Excitations},\hspace{3pt} Physical Review Letters, 50, 1395-1398, 1983. 
\bibitem{Zh1}Shou-Cheng Zhang, \hspace{3pt}\emph{The Chern-Simons-Landau-Ginzburg theory of the fractional quantum Hall
effect},\hspace{3pt}International Journal of Modern Physics, 6B, 25-58, 1992.  
\bibitem{St1}Michael Stone,\hspace{3pt} \emph{Schur functions, chiral bosons and the quantum-Hall-effect edge states},\hspace{3pt} Physical Review, 42B, 8399-8404, 1990.
\bibitem{Hu1}Jiang-Ping Hu and Shou-Cheng Zhang, \hspace{3pt}\emph{A Four-Dimensional Generalization of the quantum Hall Effect},\hspace{3pt} Science 294(5543), 823-828, 2001. 
\bibitem{Hu2}Jiang-Ping Hu and Shou-Cheng Zhang, \hspace{3pt}\emph{Collective excitations at the boundary of a 4D quantum Hall
droplet},\hspace{3pt}cond-mat/0112432.
\bibitem{Zh2}Shou-Cheng Zhang,\hspace{3pt} \emph{ To see a world in a grain of sand},\hspace{3pt} hep-th/0210162.
\eject\noindent
\bibitem{Zh3}Shou-Cheng Zhang, \hspace{3pt} \emph{Exact microscopic wave function for a topological quantum membrane},\hspace{3pt}  cond-mat/0210604.
\bibitem{Be1}B. Andrei Bernevig, Chyh-Hong Chern, Jiang-Ping Hu, Nicolaos Toumbas and Shou-Cheng Zhang,\hspace{3pt}  \emph{Effective Field Theory description of the higher dimensional
quantum Hall liquid},\hspace{3pt}  Annals of Physics, 300, 185, 2002.
\bibitem{Ch1}Yi-Xin Chen,\hspace{3pt} \emph{Matrix models of 4-dimensional quantum Hall fluids},\hspace{3pt}  hep-th/0209182.
\bibitem{Ch2}Yi-Xin\hspace{-2pt} Chen, \hspace{3pt} \emph{Quasi-particle excitations and hierarchies of 4-dimensional quantum Hall fluid states in the matrix models},\hspace{3pt} 
hep-th/0210059.
\bibitem{Ch3}Yi-Xin Chen, Bo-Yu Hou, Bo-Yuan Hou,\hspace{3pt}  \emph{Non-commutative geometry of 4-dimensional quantum Hall droplet}, \hspace{3pt} Nuclear Physics, B638, 220-242, 2002.
\bibitem{El1}Henriette Elvang and Joseph Polchinski, \hspace{3pt} \emph{The Quantum Hall Effect on $\R^4$}, \hspace{3pt} hep-th/0209104.
\bibitem{Fa1}Michal Fabinger,\hspace{3pt}\emph{Higher-Dimensional Quantum Hall Effect in String Theory}, \hspace{3pt} Journal of High Energy Physics, 0205, 037-049, 2002.
\bibitem{Ka3}Dimitra Karabali and V. Parameswaran Nair,\hspace{3pt} \emph{ Quantum Hall Effect in Higher Dimensions},\hspace{3pt}  hep-th/0203264.
\bibitem{Ki1}Yusuke Kimura, \hspace{3pt} \emph{Non-commutative Gauge Theory on Fuzzy Four-Sphere and Matrix Model}, \hspace{3pt} Nuclear Physics, B637,
177-198, 2002.
\bibitem{Sp1}George A.J. Sparling,\hspace{3pt}
\emph{Zitterbewegung}, \hspace{3pt}Seminaires et Congres, Societe Mathematique de
France, 4, 275-303, 2000.
\bibitem{Ka1}Devendra Kapadia  and George A.J. Sparling,\hspace{3pt}\emph{Glitch metrics}, \hspace{3pt}Laboratory of Axiomatics preprint,
2002.
\bibitem{Ka2}Devendra Kapadia  and George A.J. Sparling, \hspace{3pt}\emph{A class of conformally Einstein metrics}, \hspace{3pt}Classical and Quantum
Gravity, 24, 4765-4776, 2000.
\bibitem{Sp2}George A.J. Sparling, \hspace{3pt} \emph{Twistor theory and the four-dimensional Quantum Hall effect of Zhang and Hu},\hspace{3pt}cond-mat/0211679. 
 \bibitem{Ti1}George A.J. Sparling and Philip Tillman,  \hspace{3pt} \emph{A primordial theory}, \hspace{3pt}cond-mat/0401015. 
\bibitem{Su1}Leonard Susskind,\hspace{3pt}  \emph{The Quantum Hall Fluid and Non-Commutative Chern-Simons Theory}, \hspace{3pt}  hep-th/0101029, 2001.  
\bibitem{He2} Simeon Hellerman and Leonard Susskind, \hspace{3pt}\emph{Realizing the Quantum Hall System in String Theory},\hspace{3pt} hep-th/0107200.
\bibitem{Po1} Alexios P. Polychronakos,\hspace{3pt} \emph{Noncommutative Chern-Simons terms and the noncommutative vacuum}, \hspace{3pt}Journal of High Energy Physics, 0011, 008, 2000.   
\bibitem{Po2} Alexios P. Polychronakos, \hspace{3pt}\emph{Quantum Hall States as a Matrix
Chern-Simons Theory}, \hspace{3pt}hep-th/0103013.
\bibitem{Po3} Alexios P. Polychronakos, \hspace{3pt}\emph{Flux tube Solutions in Noncommutative
Gauge Theories},\hspace{3pt} hep-th/0007043.
\bibitem{Ga1} Hugo Garc\'{i}a-Compe\'{a}n, Jerzy F. Pleba\'{n}ski and Maciej
Przanowski, \hspace{3pt}\emph{The Geometry of Deformation Quantization and Self-Dual
Gravity},\hspace{3pt} hep-th/9710154.
\bibitem{St2} Ian A.B. Strachan, \hspace{3pt}\emph{The Moyal Bracket and the
Dispersionless limit of the KP Hierarchy},\hspace{3pt} hep-th/9410048.
\bibitem{Ne1} Ezra T. Newman, \hspace{3pt}\emph{Heaven and its properties},\hspace{3pt} General
Relativity and Gravitation, 7, 107-127, 1976.
\bibitem{Pe1} Roger Penrose,\hspace{3pt} \emph{Non-linear gravitons and curved twistor
theory},\hspace{3pt} General Relativity and Gravitation, 7, 31-52, 1976.
\bibitem{Ha2}Richard O. Hansen, Ezra T. Newman, Roger Penrose and K. Paul Tod, \hspace{3pt}\emph{The Metric and Curvature Properties of H-Space},\hspace{3pt} Proceedings of the Royal Society of London,  A363, 445-468, 1978.
\bibitem{Wa1} Richard S. Ward, \hspace{3pt}\emph{On self-dual gauge fields},\hspace{3pt} Physics Letters, 61A, 81-83, 1977.
\bibitem{At1}Michael F. Atiyah and Richard S. Ward,\hspace{3pt}\emph{Instantons and Algebraic Geometry}, \hspace{3pt} Communications in Mathematical Physics, 55, 117-124, 1977.
\bibitem{Pe2}Roger Penrose and Wolfgang Rindler,\hspace{3pt} \emph{ Spinors and space-time Volume 1: Two-spinor calculus and relativistic
fields},
\hspace{3pt} Cambridge University Press, Cambridge, 1984.
\bibitem{Pe3}Roger Penrose and Wolfgang Rindler,\hspace{3pt} \emph{ Spinors and space-time Volume 2: Spinor and twistor methods in space-time
geometry},
\hspace{3pt} Cambridge University Press, Cambridge, 1986.
\bibitem{Ne2}Ezra T. Newman and Roger Penrose, \hspace{3pt}\emph{An approach to gravitational radiation by a method of spin coefficients}, \hspace{3pt} Journal of
Mathematical Physics, 3, 566-578, 1962.
\bibitem{Sp3}George A.J. Sparling and K. Paul Tod,
\hspace{3pt}\emph{An Example of an H-Space}, Journal of Mathematical Physics 22, 331-332, 1981.
\bibitem{Ma1} Lionel J. Mason and George A.J. Sparling, \hspace{3pt}\emph{Non-linear Schrodinger and Korteweg-de Vries are reductions of self-dual Yang-Mills}, \hspace{3pt} Physics Letters, A137, 29-33, 1989.
\bibitem{Ma2} Lionel J. Mason and George A.J. Sparling, \hspace{3pt}\emph{ Twistor correspondences for the soliton hierarchies},\hspace{3pt} Journal of Geometry and Physics, 8, 243-271, 1992.
\bibitem{Co1} Alain Connes, \hspace{3pt}\emph{Noncommutative Geometry}, \hspace{3pt}Academic
Press, 1994.
\bibitem{Cu1} Joachim Cuntz and Daniel Quillen, \hspace{3pt}\emph{Algebraic Extensions and
Nonsingularity},\hspace{3pt} Journal of the American Mathematical Society, 8, 2, 1995.
\bibitem{As1} Abhay Ashtekar, Christopher Beetle, Olaf Dreyer, Stephen Fairhurst, Badri Krishnan, Jerzy Lewandowski and Jacek Wisniewski, \hspace{3pt}\emph{ Generic Isolated Horizons and their Applications}, \hspace{3pt}Physical Review Letters, 85, 3564-3567, 2000.
\bibitem{Be2} Jacob D. Bekenstein, \hspace{3pt}\emph{Black Holes and the Second Law}, \hspace{3pt} Lettere al Nuovo Cimento, 4, 737, 1972.
\bibitem{Ha3}  Stephen W.  Hawking, \hspace{3pt}\emph{Black Hole Explosions},\hspace{3pt}Nature, 248, 30, 1974.
\bibitem{Ha4}James M. Bardeen, Brandon Carter and Stephen W. Hawking,\hspace{3pt} \emph{The Four Laws of Black Hole Mechanics}, \hspace{3pt} Communications in Mathematical Physics, 31, 161-170, 1973.
\bibitem{Ho1} Gerard Ôt Hooft, \hspace{3pt}\emph{The Holographic Principle, Opening Lecture}, \hspace{3pt}in\hspace{3pt}\emph{Basics and Highlights in Fundamental Physics}, \hspace{3pt}The Subnuclear series, Volume 37, pages 72-100, World Scientific, 2001 (Erice, August 1999), editor Antonino Zichichi.
\bibitem{Su2} Leonard Susskind, \hspace{3pt}\emph{The World as a Hologram}, \hspace{3pt}Journal of Mathematical Physics, 36 , 6377-6396, 1995.
\bibitem{Gi1} Gary W. Gibbons, Stephen W. Hawking and Malcolm J. Perry, \hspace{3pt}\emph{Path Integrals and the Indefiniteness of the Gravitational Action},\hspace{3pt} Nuclear Physics B138, 141-150, 1978.


\end{thebibliography}
\end{document}